\documentclass[twocolumn,floatfix,eqsecnum, superscriptaddress,pra,aps]{revtex4-2}
\usepackage{graphicx}
\usepackage{amsmath}
\usepackage{amssymb}
\usepackage{bm}
\usepackage{hyperref}

\usepackage{color}

\begin{document}

\title{Fragile single-cone Dirac quantum walks in two dimensions}

\author{C. W. J. Beenakker}
\affiliation{Instituut-Lorentz, Universiteit Leiden, P.O. Box 9506, 2300 RA Leiden, The Netherlands}
\author{J. S\'{a}nchez Fern\'{a}n}
\affiliation{Instituut-Lorentz, Universiteit Leiden, P.O. Box 9506, 2300 RA Leiden, The Netherlands}
\author{J. Tworzyd{\l}o}
\affiliation{Faculty of Physics, University of Warsaw, ul.\ Pasteura 5, 02--093 Warszawa, Poland}

\date{August 2026}

\begin{abstract}
It is known that a one-dimensional (1D) quantum walk gives a \textit{local} space-time discretization of the massless Dirac equation with a \textit{single} quasi-energy cone (no fermion doubling at low energies), keeping the fundamental symmetries (chiral and time-reversal) of the continuum theory. We show that the analogous 2D construction is fundamentally more fragile. Local two-band quantum walks can have an unpaired Dirac cone, but the protecting symmetries then cease to be ordinary on-site symmetries: They become non-symmorphic, involving half-lattice translations, and are broken by generic spatial inhomogeneities. In particular, we demonstrate that the 2D Dirac quantum walk based on the Ho-Chalker network model can be gapped by potential scattering, as is evidenced by the breakdown of reflectionless Klein tunneling. The symmorphic symmetry obstruction is sharp: If we allow for couplings that are no longer strictly finite-range but decay exponentially with distance, we can construct a 2D quantum walk with a single Dirac cone and \textit{on-site} chiral and time-reversal symmetries.
\end{abstract}
\maketitle

\section{Introduction}

The connection between quantum walks (QW) and relativistic quantum dynamics can be traced back to Feynman's checkerboard construction for the one-dimensional (1D) Dirac equation \cite{Fey47,Fey65}. A discrete-time quantum walk is a local unitary update of a single-particle state on a space-time lattice: a conditional translation (the shift) composed with a unitary rotation of a spin-1/2 degree of freedom (the coin). The Dirac equation results in the continuum limit \cite{Nar72,Suc93,Bia94,Mey96,Kul99,Str06,Sun12,Arr14,Sko22}.

The emergence of massless Dirac fermions in condensed matter, in graphene and topological insulators, has called attention to the role of disorder in Dirac quantum walks \cite{Kit10,Kit12,Ahl11,Arr18,Asb20,Yam23,Pan26}. The 2D surface state of a topological insulator has a topologically protected Dirac cone \cite{Has10}, which cannot be gapped by disorder that preserves either time-reversal symmetry or chiral symmetry. \textit{Is a 2D Dirac quantum walk topologically protected?} That is the question addressed in this work.

We focus on (2+1)-dimensional space-time discretizations of the Dirac equation that produce \textit{a single} Dirac cone in the quasi-energy--momentum $\varepsilon(\bm{k})$ dispersion relation. The absence of fermion doubling is crucial for the topological protection: If there are two Dirac points at $\varepsilon=0$ in the Brillouin zone they can be coupled by disorder and gap out, without breaking any symmetry \cite{Iad24}.

\begin{figure}[tb]
\centerline{\includegraphics[width=0.8\linewidth]{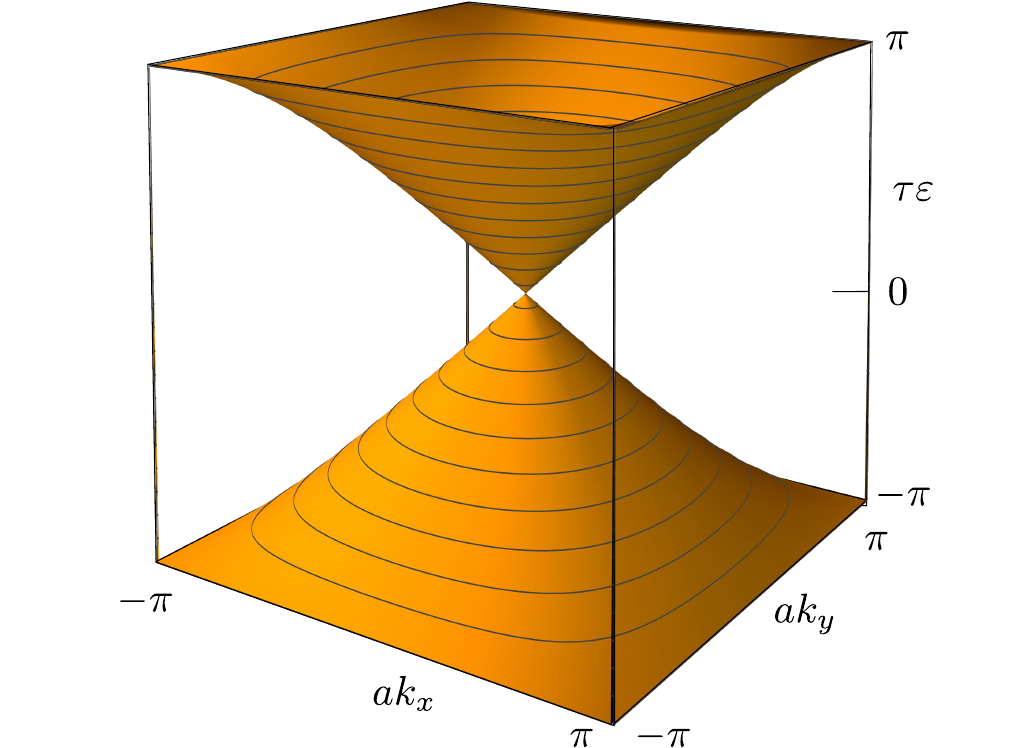}}
\caption{Momentum dependence of the quasi-energy \eqref{localdispersion} of the local 2D quantum walk considered in Sec.\ \ref{sec_squared} (squared quantum walk, Floquet operator $U_{\sqrt{XYXY}}$). Only the first Brillouin zone is shown. We study whether disorder can gap the $\varepsilon=0$ Dirac point.
}
\label{fig_cone}
\end{figure}

An example of such a single-cone dispersion that we will discuss is
\begin{equation}
\cos^2(\tau\varepsilon/2)=\cos^2(ak_x/2)\cos^2(ak_y/2),\label{localdispersion}
\end{equation}
where $a$ and $\tau$ are the space and time discretization units (see Fig.\ \ref{fig_cone}). The Dirac cone is robust in the presence of a 1D potential that varies only along the $x$ or $y$-direction, but it is gapped by a 2D potential such as $V\cos[\pi (x+y)/2a]$, showing the essential difference between 1D and 2D Dirac quantum walks. We trace the absence of topological protection in 2D to a fragile non-symmorphic symmetry \cite{Moc20,McC23}, which is broken by spatial inhomogeneities.

We will argue that the breakdown of topological protection is generic for local, two-band, single-cone quantum walks in (2+1)-dimensions: chiral and time-reversal symmetries are either broken explicitly (for example, in the twisted Dirac quantum walk \cite{Jol23,Gup25,Gup26}), or they become non-symmorphic and are broken by disorder. 

\begin{figure}[tb]
\centerline{\includegraphics[width=0.8\linewidth]{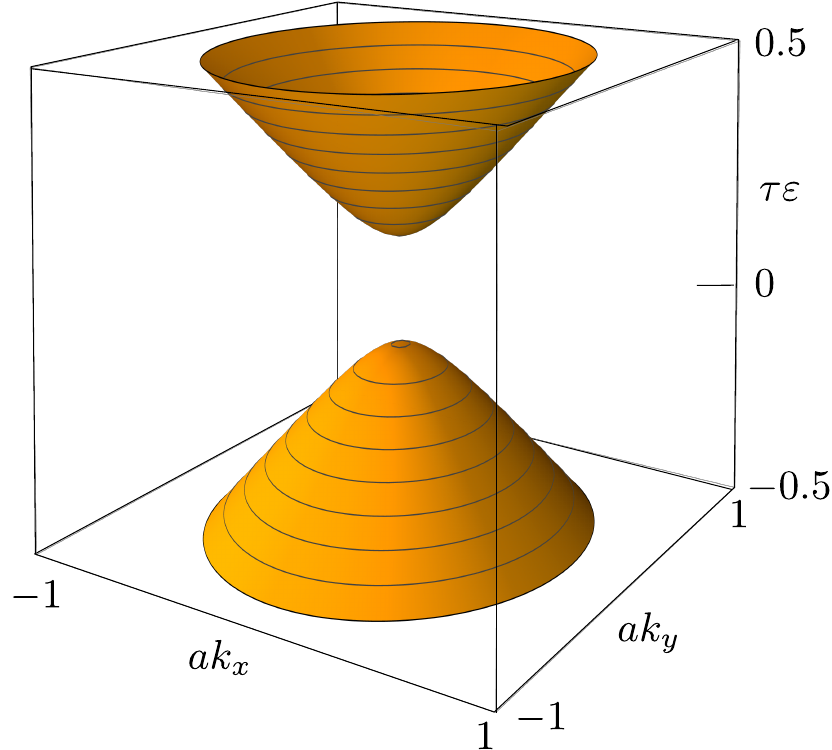}}
\caption{Gapped cone of the Ho-Chalker quantum walk in the presence of the staggered potential $V(\bm{r})=(2\tau)^{-1}\cos (\pi x/a)$.
}
\label{fig_gapped}
\end{figure}

Our roadmap is as follows: We first analyze three representative quantum walks, in Sec.\ \ref{sec_DiracQW}, giving special attention to the Dirac quantum walk \cite{Lia13,Pas14,Del17} obtained from the Ho-Chalker network model \cite{Ho96}. In Sec.\ \ref{sec_stability} we show that it is gapped by a staggered potential (see Fig.\ \ref{fig_gapped}), because of the fragility of the non-symmorphic symmetries. To demonstrate that this is generic, we prove a no-go argument in Sec.\ \ref{sec_nogo}, highlighting why it does not apply to 1D quantum walks. In Sec.\ \ref{sec_Klein} we make contact with a measurable quantity, by examining Klein tunneling through a potential barrier (Fig.\ \ref{fig_klein}). Finally, we investigate how much of the locality has to be given up to restore the protection. Reversing the mapping that underlies our no-go argument, we obtain in Sec.\ \ref{sec_exploc} a single-cone walk with genuine on-site symmetries from a flattened Chern band, with hopping amplitudes that decay \textit{exponentially} over a few lattice sites.

\section{Dirac quantum walks}
\label{sec_DiracQW}

We consider a cubic (2+1)-dimensional space-time lattice (lattice spacing $a$ in space and $\tau$ in time), on which a single-particle state evolves unitarily,
\begin{equation}
\psi(t+\tau)=F\psi(t),\;\;FF^\dagger=\openone.
\end{equation}
The unitary Floquet operator $F$ acts on the position $|\bm{r}\rangle$ ($\bm{r}\in\mathbb{Z}^2$) of the particle and on its internal spin-degree of freedom $|s\rangle$, $s\in\{\uparrow,\downarrow\}$. If $F$ acts locally, only coupling nearby lattice sites, this stroboscopic dynamics is referred to as a quantum walk. The spin represents the ``coin'' that is flipped by $F$, jointly with a spin-dependent displacement. The eigenvalues $e^{-i\varepsilon_n\tau/\hbar}$ of $F$ give the quasi-energies $\varepsilon_n$, defined modulo $2\pi\hbar/\tau$. We set $\hbar$ to unity, so the (2+1)-dimensional Brillouin zone is $ak_x,ak_y,\tau\varepsilon\in(-\pi,\pi]$.

A Dirac quantum walk is a quantum walk which in the continuum limit reduces to the Dirac equation,
\begin{equation}
\begin{split}
&i\partial_t\psi=H_{\rm Dirac}\psi,\\
&H_{\rm Dirac}=v_{\rm eff}(p_x\sigma_x+p_y\sigma_y)+V\sigma_0+\bm{M}\cdot\bm{\sigma}.
\end{split}
\label{HDirac}
\end{equation}
The momentum operator is $p_\alpha=-i\partial/\partial x_\alpha\equiv -i\partial_\alpha$, the Pauli matrices $\bm{\sigma}=(\sigma_x,\sigma_y,\sigma_z)$ act on the spin degree of freedom (with $\sigma_0$ the $2\times 2$ unit matrix). The effective velocity $v_{\rm eff}$ may or may not differ from the ratio $v=a/\tau$. The scalar potential $V(\bm{r})$ and magnetization vector $\bm{M}(\bm{r})$ are position dependent.

\subsection{Palindromic quantum walk}

The simplest 2D example \cite{Arr14} takes the product of Feynman's 1D propagator,
\begin{equation}
U_{XY} = K_x K_y,\;\;K_\alpha=e^{-iap_\alpha\sigma_\alpha}.\label{UXY}
\end{equation}
Following Asb\'{o}th \cite{Asb12,Asb13}, the palindromic order
\begin{equation}
U_{\sqrt{XYYX}}=K_x^{1/2}K_y K_x^{1/2}\label{UsqrtXYYX}
\end{equation}
is preferred: Both unitaries \eqref{UXY} and \eqref{UsqrtXYYX} have the same spectrum, but the second satisfies the chiral symmetry relation $\sigma_z U\sigma_z=U^{-1}$, while the first does not.

Including also the potentials $V,\bm{M}$ in palindromic order, we have the Floquet operator
\begin{equation}
\begin{split}
&F_{\sqrt{XYYX}}=U_{VM}U_{\sqrt{XYYX}}U_{VM},\\
&U_{VM}=e^{-i(V\sigma_0+\bm{M}\cdot\bm{\sigma})\tau/2}.
\end{split}\label{FXYYX}
\end{equation}

In terms of the unitary shift operators,
\begin{equation}
\begin{split}
&T_\alpha=e^{iap_\alpha}=e^{a\partial_\alpha},\\
&T_x|x,y\rangle=|x+1,y\rangle,\;\;T_y|x,y\rangle=|x,y+1\rangle,
\end{split}
\end{equation}
one has
\begin{align}
&U_{\sqrt{XYYX}}=\tfrac{1}{4}(T_x+T_x^{-1}) (T_y+T_y^{-1})\sigma_0\nonumber\\
&\;\;-\tfrac{1}{4}(T_x-T_x^{-1}) (T_y+T_y^{-1})\sigma_x-\tfrac{1}{2}(T_y-T_y^{-1})\sigma_y.
\end{align}
Note that only integer shifts appear in $U_{\sqrt{XYYX}}$, although $K^{1/2}_x$ is a half-integer shift.

The eigenvalues of $T_\alpha$ are $e^{iak_\alpha}$, $k_\alpha\in(-\pi/a,\pi/a]$. The resulting eigenvalues of $U_{\sqrt{XYYX}}$ give the quasi-energy-momentum dispersion relation \cite{note1}
\begin{equation}
\cos\tau\varepsilon=\cos (ak_x)\cos (ak_y)\;\;
\text{[palindromic QW]}.
\end{equation}
Near $(k_x,k_y)=(0,0)$ this expands to the Dirac cone of the continuum,
\begin{equation}
\varepsilon^2=v^2(k_x^2+k_y^2)+{\cal O}(k^4),\;\;v=a/\tau.
\end{equation}
There is a second, spurious, zero-quasi-energy Dirac cone at the corner $(k_x,k_y)=(\pi/a,\pi/a)$ of the Brillouin zone. This is the fermion doubling deficiency of lattice fermions, which one seeks to avoid.

\subsection{Squared quantum walk}
\label{sec_squared}

An alternative quantum walk squares the operator $K_x^{1/2}K_y^{1/2}$,
\begin{equation}
\begin{split}
&F_{\sqrt{XYXY}}=U_{VM}U_{\sqrt{XYXY}}U_{VM},\\
&U_{\sqrt{XYXY}}=K_x^{1/2}K_y^{1/2}K_x^{1/2}K_y^{1/2},
\end{split}
\label{FXYXY}
\end{equation}
with dispersion relation
\begin{align}
&\cos\tau\varepsilon=\tfrac{1}{2}\bigl[\cos ak_x+\cos ak_y+\cos (ak_x)\cos (a k_y)-1\bigr]\nonumber\\
&\Rightarrow\cos^2(\tau\varepsilon/2)=\cos^2(ak_x/2)\cos^2(ak_y/2)\\
&\text{[squared QW]}.\nonumber
\end{align}
The expansion around $(k_x,k_y)=(0,0)$ is again a Dirac cone with velocity $v=a/\tau$.

Note the close structural similarity between $U_{\sqrt{XYYX}}$ [Eq.\ \eqref{UsqrtXYYX}] and $U_{\sqrt{XYXY}}$: the only difference is in the order by which the $x$ and $y$ steps are applied. This change has a qualitative effect on the dispersion relation: The doubler at $\varepsilon=0$, $(k_x,k_y)=(\pi/a,\pi/a)$ has been removed, producing the unpaired Dirac cone shown in Fig.\ \ref{fig_cone}. Notice also that the squared QW has $\varepsilon=\pi/\tau$ along the entire momentum boundary of the Brillouin zone, while for the palindromic QW $\varepsilon=\pi/\tau$ at isolated momenta [$(k_x,k_y)=(\pi/a,0)$ and $(0,\pi/a)$].

\subsection{Ho-Chalker quantum walk}

The Ho-Chalker network model \cite{Ho96} gives another way to remove the doubler. Its Floquet operator is \cite{Del17}
\begin{equation}
F_{\text{Ho-Chalker}}=U_{VM}U_{\text{Ho-Chalker}}U_{VM},\label{FHC}
\end{equation}
with $U_{VM}$ as defined in Eq.\ \eqref{FXYYX} and \cite{note3}
\begin{align}
&U_{\text{Ho-Chalker}}=e^{-ia(p_x+p_y)\sigma_x/2}e^{-ia(p_x-p_y)\sigma_y/2}\label{UHC}\\
&=\frac{1}{\sqrt{2}}e^{-i(\pi/4)\sigma_x}e^{-i(\pi/4)\sigma_y}\begin{pmatrix}
T_y^{-1} & T_x^{-1}\\
-T_x& T_y\\
\end{pmatrix}e^{i(\pi/4)\sigma_x}.\nonumber
\end{align}
The dispersion relation now reads
\begin{equation}
\cos\tau\varepsilon=\tfrac{1}{2}\cos ak_x+\tfrac{1}{2}\cos ak_y\;\;
\text{[Ho-Chalker QW]}.
\end{equation}
Near $(k_x,k_y)=(0,0)$ this expands again to a Dirac cone (with effective velocity $v_{\rm eff}=2^{-1/2}a/\tau$), and this is the only $\varepsilon=0$ Dirac point.

\subsection{Non-symmorphic symmetries}

For $V=0=M$ the Dirac Hamiltonian \eqref{HDirac} anticommutes with the $\sigma_z$ Pauli matrix. As noted, this chiral symmetry is inherited by the palindromic quantum walk, in the form
\begin{equation}
\sigma_z U_{\sqrt{XYYX}}\sigma_z=K_x^{-1/2}K_y^{-1/2}K_y^{-1/2}K_x^{-1/2}=U_{\sqrt{XYYX}}^{-1}.
\end{equation}
For $U_{\sqrt{XYXY}}$ the chiral symmetry is non-symmorphic, meaning that it is accompanied by a half-lattice-spacing translation,
\begin{align}
\sigma_z U_{\sqrt{XYXY}}\sigma_z={}&K_x^{-1/2}K_y^{-1/2}K_x^{-1/2}K_y^{-1/2}\nonumber\\
={}&K_y^{1/2}U_{\sqrt{XYXY}}^{-1}K_y^{-1/2}.
\end{align}

Similarly, the time-reversal symmetry of $H_{\rm Dirac}=\sigma_y H_{\rm Dirac}^\ast\sigma_y$, for $V=0=M$, applies directly to $U_{\sqrt{XYYX}}$,
\begin{equation}
\sigma_y U_{\sqrt{XYYX}}^\ast\sigma_y=U_{\sqrt{XYYX}}^{-1},
\end{equation}
while for $U_{\sqrt{XYXY}}$ one has
\begin{equation}
\sigma_y U_{\sqrt{XYXY}}^\ast\sigma_y=K_y^{1/2}U_{\sqrt{XYXY}}^{-1}K_y^{-1/2}.
\end{equation}
The same applies to $U_{\text{Ho-Chalker}}$, upon replacement of conjugation by $K_y^{1/2}$ with conjugation by $e^{ia(p_x-p_y)\sigma_y/2}$.

The distinction between a symmorphic and a non-symmorphic symmetry is crucial in the presence of spatial inhomogeneities: For $F_{\sqrt{XYYX}}$ the chiral symmetry is preserved by an in-plane magnetization $\bm{M}=(M_x,M_y,0)$, while time-reversal symmetry is broken. It's the other way around for a scalar potential, which preserves time-reversal symmetry of $F_{\sqrt{XYYX}}$ while breaking chiral symmetry. In contrast, for $F_{\sqrt{XYXY}}$ and $F_{\text{Ho-Chalker}}$ the non-symmorphic symmetries are in general broken by spatially dependent $V$ or $\bm{M}$, since $U_{VM}$ does not in general commute with the half-lattice-spacing translations.

\section{Stability of the Dirac point}
\label{sec_stability}

To test for topological protection of the gapless state at the $\varepsilon=0$, $\bm{k}=0$ Dirac point, we apply the cosine potential $V(\bm{r})=V\cos(\bm{Q}\cdot\bm{r}/a)$. The calculation is outlined in App.\ \ref{app_bandstructure}. 

The choice $\bm{Q}=(\pi ,0)$ or $(0,\pi )$ or $(\pi ,\pi )$ is a staggered potential, alternating sign on neighboring lattice sites in one direction (along $x$, along $y$, or diagonally). The gap $\Delta$ at the Dirac point is then entirely determined by the unperturbed quasi-energy $\varepsilon_{\bm{Q}}$ at $\bm{k}=\bm{Q}/a$, $V=0$,
\begin{equation}
\Delta=(2/\tau)\arccos\left[\cos( \tau V)\cos(\tfrac{1}{2}\tau\varepsilon_{\bm{Q}})\right]-\varepsilon_{\bm{Q}}.\label{Egap}
\end{equation}
We see that the Dirac point remains gapless, $\Delta=0$, if and only if $\varepsilon_{\bm{Q}}=\pi/\tau$. 

As a consequence, the palindromic quantum walk remains gapless for $\bm{Q}=(\pi ,0)$ and $(0,\pi )$, but becomes gapped for $\bm{Q}=(\pi ,\pi )$, while for the Ho-Chalker quantum walk it's the other way around: gapless for $\bm{Q}=(\pi ,\pi)$ but gapped for $(\pi ,0)$ and $(0,\pi )$ (see Fig.\ \ref{fig_gapped}). The absence of fermion doubling for the Ho-Chalker quantum walk does not protect the Dirac cone, but the induced gap is of higher order in $V$,
\begin{equation}
\Delta=\begin{cases}
2 V&\text{palindromic QW},\;\;\bm{Q}=(\pi ,\pi),\\
\tau V^2+{\cal O}(V^4)&\text{Ho-Chalker QW},\;\;\bm{Q}=(\pi,0).
\end{cases}
\label{DeltaHC}
\end{equation}

\begin{figure}[tb]
\centerline{\includegraphics[width=0.8\linewidth]{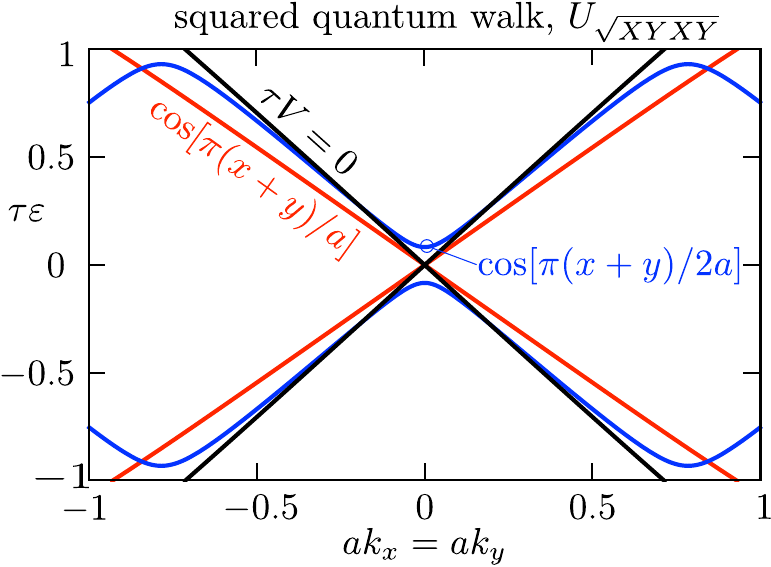}}
\caption{Cut along the line $k_x=k_y$ through the band structure of the squared quantum walk, in the presence of a scalar potential $V(\bm{r})=V\cos(\bm{Q}\cdot\bm{r}/a)$: the black curve is without any potential ($V=0$), the red and blue curves are for $\tau V=1$ and, respectively, $\bm{Q}=(\pi,\pi)$ and $(\pi/2,\pi/2)$. The latter choice gaps the $\varepsilon=0$ Dirac cone.
}
\label{fig_squared}
\end{figure}

The squared quantum walk, in contrast, remains gapless under a staggered potential in any direction, because it has $\varepsilon_{\bm{Q}}=\pi/\tau$ along the entire momentum boundary of the Brillouin zone. Still, the improved robustness does not extend to cosine potentials with a longer period (see Fig.\ \ref{fig_squared}): If we take $\bm{Q}=(\pi /2,\pi/2 )$ a gap opens also for the squared quantum walk, of magnitude
\begin{equation}
\Delta=\tfrac{1}{6}\tau V^2+{\cal O}(V^4)\;\;\text{squared QW},\;\;\bm{Q}=(\pi/2,\pi/2).
\end{equation}

\begin{figure}[tb]
\centerline{\includegraphics[width=0.8\linewidth]{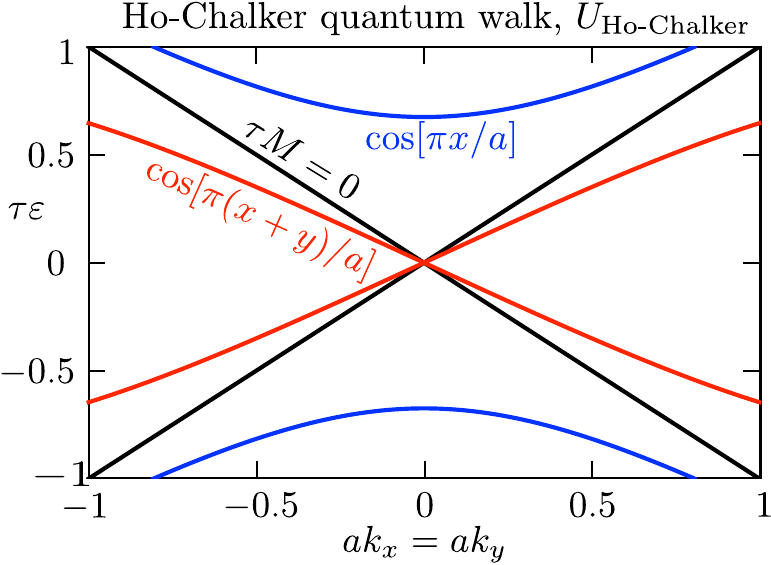}}
\caption{Cut along the line $k_x=k_y$ through the band structure of the Ho-Chalker quantum walk, in the presence of an in-plane magnetization $\bm{M}(\bm{r})=(M,M,0)\cos(\bm{Q}\cdot\bm{r}/a)$: the black curve is without any magnetization $(M=0)$, the red and blue curves are for $\tau M=1$ and, respectively, $\bm{Q}=(\pi,\pi)$ and $(\pi,0)$. The latter choice gaps the $\varepsilon=0$ Dirac cone.
}
\label{fig_HC}
\end{figure}

Scattering by the scalar potential $V$ preserves time-reversal symmetry of the Dirac Hamiltonian, while breaking chiral symmetry. We have also studied the converse: scattering by an in-plane magnetization $\bm{M}=(M_x,M_y,0)$, which preserves chiral symmetry while breaking time-reversal symmetry. The results are very similar to those for potential scattering, see Fig.\ \ref{fig_HC}.

\section{Fundamental difference between 1D and 2D quantum walks}
\label{sec_nogo}

Topological protection of the $\varepsilon=0$ Dirac point requires (1) an unpaired Dirac cone and (2) a symmorphic chiral or time-reversal symmetry. The 1D quantum walk satisfies both requirements with a local Floquet operator $U=e^{-a\partial_x\sigma_x}$: a single-cone dispersion, $\cos\tau\varepsilon =\cos a k_x $, and symmorphic chiral and time-reversal symmetries, $\sigma_z U\sigma_z=U^{-1}$, $\sigma_y U^\ast\sigma_y=U^{-1}$. The periodicity of the quasi-energy works around the Nielsen-Ninomiya fermion-doubling obstruction for local Hamiltonians \cite{Nie81}.

For the three quantum walks studied so far, the palindromic quantum walk satisfies (2) and violates (1), while the squared and Ho-Chalker quantum walks satisfy (1) and violate (2). We would now like to show this is generic for 2D Dirac quantum walks: (1) and (2) cannot both be satisfied by a local two-band Floquet matrix.

\subsection{Chiral symmetry}
\label{sec_chiral}

We first consider the case of a 2D quantum walk with a symmorphic \textit{chiral} symmetry. The Floquet matrix $U$ is an element of ${\rm SU}(2)$: chirality plus unitarity enforces $\det U=\pm  1$ and $U=\sigma_0$ at the Dirac point selects the determinant-one sector. (Since $U(\bm{k})$ is a continuous function of $\bm{k}$ in a connected Brillouin zone, $\det U=1$ at one point selects $\det U=1$ everywhere.)

The chiral symmetry $\sigma_z U\sigma_z=U^\dagger$ gives the parameterization
\begin{equation}
    U=a\sigma_0+i b\sigma_x+i c\sigma_y+d\sigma_z,
    \;\; a,b,c,d\in{\mathbb R},
\end{equation}
and unitarity gives
\begin{equation}
    a^2+b^2+c^2+d^2=1,\;\; ad=bd=cd=0.
\end{equation}
The branch with $d\neq 0$ has $U=\pm\sigma_z$ and determinant $-1$.
Hence in the determinant-one sector one has $d=0$.

The Floquet operator then has the form
\begin{equation}
 U(\bm{k})
        =
        n_0(\bm{k})\sigma_0
        +i n_x(\bm{k})\sigma_x
        +i n_y(\bm{k})\sigma_y ,
\end{equation}
for a vector field $\bm{n}=(n_x,n_y,n_0)$ that assigns a point on the unit sphere $\mathbb{S}^2$,
\begin{equation}
n_x^2(\bm{k})+n_y^2(\bm{k})+n_0^2(\bm{k})=1,\label{nnormalized}
\end{equation}
to each momentum $\bm{k}=(k_x,k_y)$ in the Brillouin zone (the torus $\mathbb{T}^2)$.

We require a single $\varepsilon=0$ Dirac point in the Brillouin zone, so a single $\bm{k}_\ast$ at which $U(\bm{k}_\ast)=\sigma_0\Rightarrow\bm{n}(\bm{k}_\ast)=(0,0,1)\equiv\bm{N}$. The Floquet operator thus defines a map $\bm{n}:\mathbb{T}^2\rightarrow \mathbb{S}^2$ from the torus to the sphere, and we inquire whether it is possible that $\bm{n}(\bm{k})=\bm{N}$ at a single point $\bm{k}_\ast$. Because the Dirac point is a linear band crossing, $\bm{N}$ is a regular (locally invertible) point of the map.

We need one basic concept from algebraic topology \cite{Hat01}: The degree $\operatorname{deg}\bm{n}$ of a continuous map is an integer which counts how many times the domain manifold ($\mathbb{T}^2$ in our case) wraps around the target manifold ($\mathbb{S}^2$), accounting for orientation. Each pre-image $\bm{k}_i$ of a regular point $\bm{N}$ contributes $+1$ or $-1$ to the degree, depending on whether the map preserves or reverses orientation near $\bm{k}_i$. Since we assume that $\bm{N}$ has a single pre-image $\bm{k}_\ast$, we necessarily have $\operatorname{deg}\bm{n}=\pm 1$. 

We now argue by contradiction:  Consider, instead of the unitary $U$, the Hermitian
\begin{equation}
H(\bm{k})=n_x(\bm{k})\sigma_x+n_y(\bm{k})\sigma_y+n_0(\bm{k})\sigma_z,\label{Hflat}
\end{equation}
which like $U$ is a local operator. The normalization \eqref{nnormalized} enforces $H^2=\sigma_0$, hence this fictitious Hamiltonian has a pair of exactly flat non-degenerate bands, one at eigenvalue $+1$ and the other at eigenvalue $-1$, with Chern number $\pm 1$ set by the degree of the map. We have thus constructed a non-degenerate flat-band with non-zero Chern number from a local Hamiltonian, which is forbidden \cite{Che14,Rea17}.

We conclude that a local, two-band, 2D Dirac quantum walk with a symmorphic chiral symmetry cannot have one single $\varepsilon=0$ Dirac point in the Brillouin zone. Note that the presence or absence of additional band crossings at higher quasi-energy plays no role in this argument.

To reach a contradiction, we have made essential use of the fact that we are not in 1D: a map from $\mathbb{T}^1$ to $\mathbb{S}^2$ has no regular points and no degree, so a single pre-image of $\bm{N}$ carries no obstruction --- consistent with the existence of the 1D single-cone Dirac quantum walk with chiral symmetry.

\subsection{Time-reversal symmetry}

Now we turn to the case of a 2D quantum walk with a symmorphic \textit{time-reversal} symmetry. The unitary Floquet matrix $U(\bm{k})$ in Fourier representation is constrained by
\begin{equation}
\sigma_y U^\ast(-\bm{k})\sigma_y=U^\dagger(\bm{k}).
\end{equation}
This implies that the determinant $\det U(\bm{k})=e^{i\phi(\bm{k})}$ must satisfy $\phi(\bm{k})=\phi(-\bm{k})$. Locality restricts the phase $\phi(\bm{k})$ to a linear function of $\bm{k}$, so it can only be even if $\phi(\bm{k})=\phi_0$ is a constant. Since a constant phase offset merely shifts the zero of quasi-energy, we may without loss of generality set $\phi_0=0\Rightarrow \det U(\bm{k})=1$.

The matrix $U\in {\rm SU}(2)$ is parameterized by
\begin{equation}
U(\bm{k})
        =
        n_0(\bm{k})\sigma_0
        +i n_x(\bm{k})\sigma_x
        +i n_y(\bm{k})\sigma_y 
        +i n_z(\bm{k})\sigma_z,\label{UTRS}
\end{equation}
with $n_j(\bm{k})$ a real function of $\bm{k}=(k_x,k_y)\in\mathbb{T}^2$. Unitarity and time-reversal symmetry constrain these functions by
\begin{align}
&n_0^2(\bm{k})+n_x^2(\bm{k})+n_y^2(\bm{k})+n_z^2(\bm{k})=1,\label{TRSunitarity}\\
&n_0(-\bm{k})=n_0(\bm{k}),\;\;n_\alpha(-\bm{k})=-n_\alpha(\bm{k}),\;\;\alpha\in\{x,y,z\}.\nonumber
\end{align}

Kramers degeneracy produces a band crossing at each of four time-reversally invariant momenta,
\begin{equation}
\{\bm{k}_1,\bm{k}_2,\bm{k}_3,\bm{k}_4\}=\{(0,0),(\pi/a,0),(0,\pi/a),(\pi/a,\pi/a)\},\label{TRIM}
\end{equation}
such that $\bm{k}_p=-\bm{k}_p$ on the torus $\mathbb{T}^2$. At each $\bm{k}_p$ the odd functions $n_x,n_y,n_z$ must vanish, and the normalization then requires that $n_0=\pm 1$. These are four Dirac points, at $\varepsilon=0$ if $n_0=+1$ and at $\varepsilon=\pi/\tau$ if $n_0=-1$.

If we define the parity
\begin{equation}
{\cal P}=\prod_{p=1}^4 n_0(\bm{k}_p)\in\{+1,-1\},\label{Pdef}
\end{equation}
an unpaired $\varepsilon=0$ Dirac cone requires ${\cal P}=-1$. This is forbidden by locality of $U(\bm{k})$, see App.\ \ref{app_parity} for a proof.

\section{Klein tunneling}
\label{sec_Klein}

To demonstrate the fragility of the 2D quantum walk for a measurable quantity we consider \textit{Klein tunneling}: In the continuum, a massless Dirac fermion normally incident on a potential barrier is transmitted with unit probability, irrespective of the barrier height. Reflection of the particle would require a chirality flip which is forbidden by time-reversal symmetry \cite{Kat06,Bee08}.

We study Klein tunneling of a quantum walk through a potential barrier $V(x)$ along the $y$-direction, so that $p_y=0$ is a good quantum number. Both the palindromic and squared quantum walk then reduce to Feynman's 1D propagator, $U= e^{-iap_x\sigma_x}$ for $p_y=0$. An incident plane wave $\psi(x)=e^{i\varepsilon\tau x}{1\choose 1}$ at quasi-energy $\varepsilon$ acquires a phase factor $t=\exp[-i\tau\sum_x V(x)]$ upon propagation through the barrier. The transmission probability $T=|t|^2=1$ irrespective of the potential profile.

As we will now show, this reflectionless tunneling effect breaks down for the Ho-Chalker quantum walk, which for $p_y=0$ has propagator
\begin{equation}
U_{\text{Ho-Chalker}}(p_y=0)=e^{-iap_x\sigma_x/2}e^{-iap_x\sigma_y/2}.\label{U0def}
\end{equation}
We consider a rectangular barrier $V(x)$, of height $V_0$ and $w$ lattice sites wide,
\begin{equation}
V(x)=\begin{cases}
V_0&\text{if}\;\; x\in\{1,2,\ldots w\},\\
0&\text{if}\;\;\text{otherwise}.
\end{cases}
\end{equation}
To simplify the expressions we take quasi-energy $\varepsilon=0$ (Dirac point).

Outside of the barrier region, the incident, transmitted, and reflected waves at $\varepsilon=0$, $p_y=0$ are given by
\begin{subequations}
\label{phioutsidebarrier}
\begin{align}
&\psi(x)=\begin{cases}
\chi_+ + r\chi_-&\text{if}\;\;x\leq 0,\\
t\chi_+&\text{if}\;\;x\geq w+1,
\end{cases}\\
&\chi_\pm=\tfrac{1}{2}{{1-i}\choose{\pm \sqrt 2}}.
\end{align}
\end{subequations}
The spinors $\chi_\pm$ are eigenstates of $\sigma_x+\sigma_y$, carrying unit current in the $\pm x$ direction.

\begin{figure}[tb]
\centerline{\includegraphics[width=0.8\linewidth]{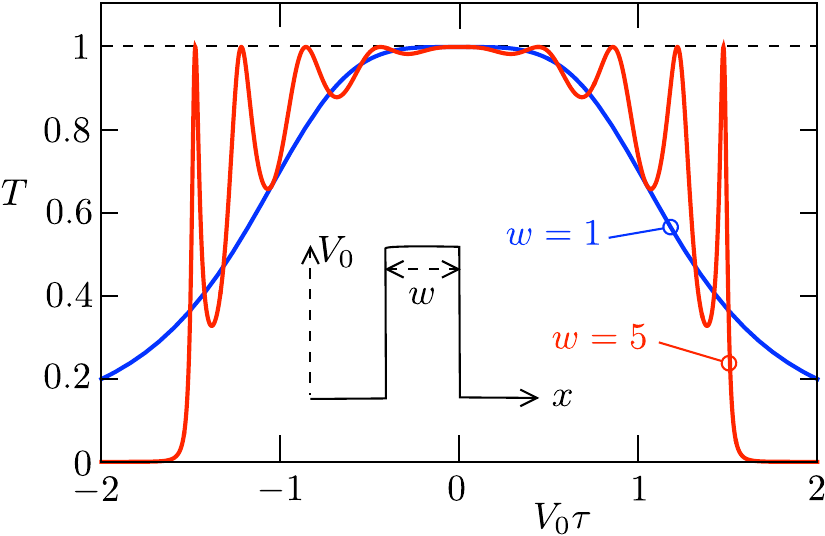}}
\caption{Transmission probability $T$ of the Ho-Chalker quantum walk through a rectangular potential barrier (height $V_0$, width $w$ lattice sites, see inset). The two plots are computed from Eq.\ \eqref{Tresult} for $w=1$ and $w=5$. The 1D quantum walk has $T=1$ for any potential height (Klein tunneling).
}
\label{fig_klein}
\end{figure}

Inside the barrier, the scattering state is the superposition of the two eigenstates of the unitary \eqref{U0def} with eigenvalue $e^{iV_0\tau}$,
\begin{subequations}
\label{phiinsidebarrier}
\begin{align}
&\psi(x)=A_+ \xi_+e^{-ik_0x} + A_- \xi_- e^{ik_0x},\;\;1\leq x\leq w,\\
\begin{split}
&\xi_\pm=\begin{pmatrix}
(1-i)\sqrt{\cos V_0\tau}\\
\pm 2\cos(V_0\tau/2-\pi/4)
\end{pmatrix},\\
&k_0=\operatorname{sign}(V_0)\arccos(-1+2\cos V_0\tau).
\end{split}
\end{align}
\end{subequations}
Mode matching, see App.\ \ref{app_Klein}, then gives the transmission and reflection amplitudes $t,r$. The resulting transmission probability is
\begin{equation}
T=\left[1+\frac{\sin^2(V_0\tau/2)}{\cos (V_0\tau)}\sin^2 (ak_0w)\right]^{-1},\label{Tresult}
\end{equation}
see Fig.\ \ref{fig_klein}

For a one-site barrier, $w=1$, this reduces to
\begin{equation}
T=\left[1+8\sin^4 (V_0\tau/2)\right]^{-1}\;\;\text{for}\;\;w=1,
\end{equation}
so a reflection probability $R=1-T\simeq \tfrac{1}{2}(V_0\tau)^4$ for a weak potential. The reflection amplitude is second order in $V_0$,  consistent with the quadratic gap opening of the Ho-Chalker quantum walk found in Eq.\ \eqref{DeltaHC}. For larger $w$ the transmission probability shows Fabry-P\'{e}rot oscillations, provided $\cos V_0\tau>0$. If $\cos V_0 \tau<0$ the wave number $k_0$ is complex and $T$ decays exponentially with increasing $w$.

\section{Exponentially local quantum walk}
\label{sec_exploc}

The no-go arguments of Sec.\ \ref{sec_nogo} assume a \textit{strictly} local Floquet operator: the hopping amplitudes vanish identically beyond a finite range. We now show that this assumption is sharp. As soon as strict locality is weakened to \textit{exponential} locality, a single unpaired $\varepsilon=0$ Dirac cone allows for on-site (symmorphic) chiral and time-reversal symmetries. We set $a=\tau=1$ in this section, so that $k_x,k_y,\varepsilon\in(-\pi,\pi]$.

\subsection{Inversion of the flat-band mapping}
\label{sec_inverting}

To construct the exponentially local quantum walk we invert the flat-band mapping of Sec.\ \ref{sec_chiral}. We start from the Qi-Wu-Zhang Hamiltonian \cite{Qi06},
\begin{align}
&H_{\rm QWZ}(\bm{k})=d_x(\bm{k})\sigma_x+d_y(\bm{k})\sigma_y+d_0(\bm{k})\sigma_z,\label{HQWZ}\\
&d_x=\sin k_x,\;\;d_y=\sin k_y,\;\;d_0=\gamma+\cos k_x+\cos k_y,\nonumber
\end{align}
which has a Chern number 1 band for $-2<\gamma<0$. We flatten the band to obtain a Hamiltonian of the form \eqref{Hflat}, with
\begin{equation}
\begin{split}
&\bm{n}=(n_x,n_y,n_0),\;\;n_\alpha(\bm{k})=d_\alpha(\bm{k})/\sqrt{D(\bm{k})},\\
&D(\bm{k})=d_x^2+d_y^2+d_0^2>0,\;\;\text{for}\;\;-2<\gamma<0.
\end{split}
\label{flattening}
\end{equation}

\begin{figure}[tb]
\centerline{\includegraphics[width=0.8\linewidth]{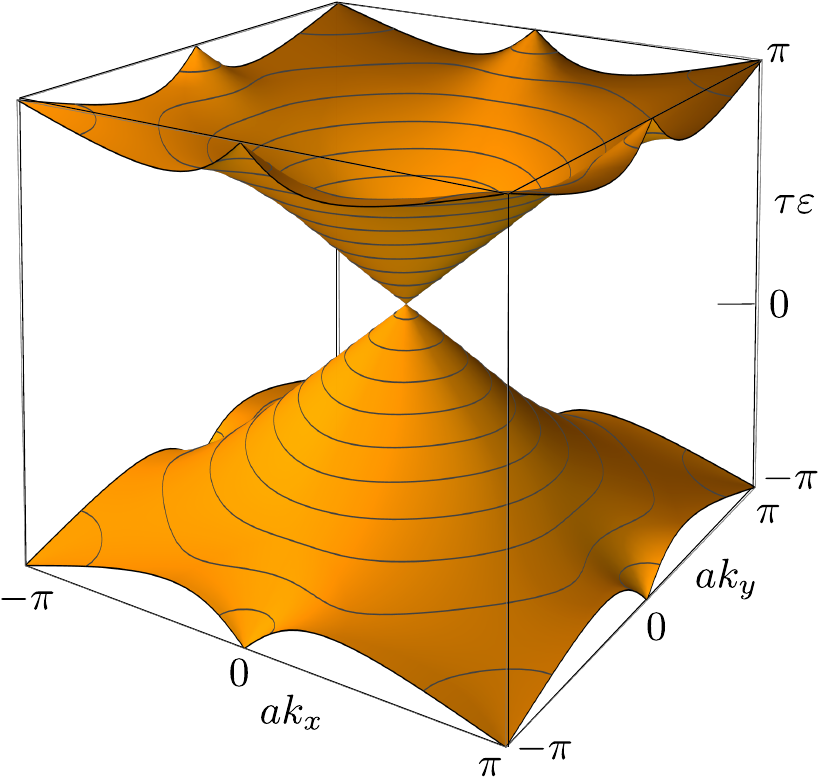}}
\caption{Quasi-energy spectrum \eqref{QWZdispersion} (plotted for $\gamma=-1$) of the 2D quantum walk with Floquet operator \eqref{UQWZ}. The single Dirac cone at $\varepsilon=0$ is protected from gap opening by symmorphic chiral and time-reversal symmetries. The quantum walk is not strictly local, but the coupling strength decays exponentially with distance.
}
\label{fig_QWZ}
\end{figure}

The corresponding unitary Floquet operator
\begin{equation}
U_{\rm QWZ}(\bm{k})=n_0(\bm{k})\sigma_0+in_x(\bm{k})\sigma_x+in_y(\bm{k})\sigma_y,
\label{UQWZ}
\end{equation}
has quasi-energy spectrum
\begin{equation}
\cos\varepsilon(\bm{k})=n_0(\bm{k})=\frac{\gamma+\cos k_x+\cos k_y}{\sqrt{D(\bm{k})}},
\label{QWZdispersion}
\end{equation}
see Fig.\ \ref{fig_QWZ}. There is a single $\varepsilon=0$ Dirac point, with effective velocity $v_{\rm eff}=(2+\gamma)^{-1}$.

Both chiral and time-reversal symmetries are on-site,
\begin{equation}
\begin{split}
&\sigma_z U_{\rm QWZ}(\bm{k})\sigma_z=U_{\rm QWZ}^\dagger(\bm{k}),\\
&\sigma_y U^\ast_{\rm QWZ}(-\bm{k})\sigma_y=U^\dagger_{\rm QWZ}(\bm{k}),
\end{split}
\label{QWZsymmetries}
\end{equation}
the second following from the parity relations $n_0(-\bm{k})=n_0(\bm{k})$, $n_{x,y}(-\bm{k})=-n_{x,y}(\bm{k})$. The no-go theorems of Sec.\ \ref{sec_nogo} are evaded, at the expense of a non-local Floquet operator --- the denominator $1/\sqrt{D(\bm{k})}$ in the definition of $\bm{n}(\bm{k})$ has infinite range in real space. In particular, the time-reversally invariant momenta \eqref{TRIM} have three Dirac points at $\varepsilon=\pi$ and one at $\varepsilon=0$, so the parity \eqref{Pdef} equals ${\cal P}=-1$. The proof in App.\ \ref{app_parity} that ${\cal P}=+1$ does not apply because it assumes that $\bm{n}(\bm{k})$ is a finite trigonometric polynomial, which $1/\sqrt{D(\bm{k})}$ is not.

\begin{figure*}[tb]
\centerline{\includegraphics[width=0.8\linewidth]{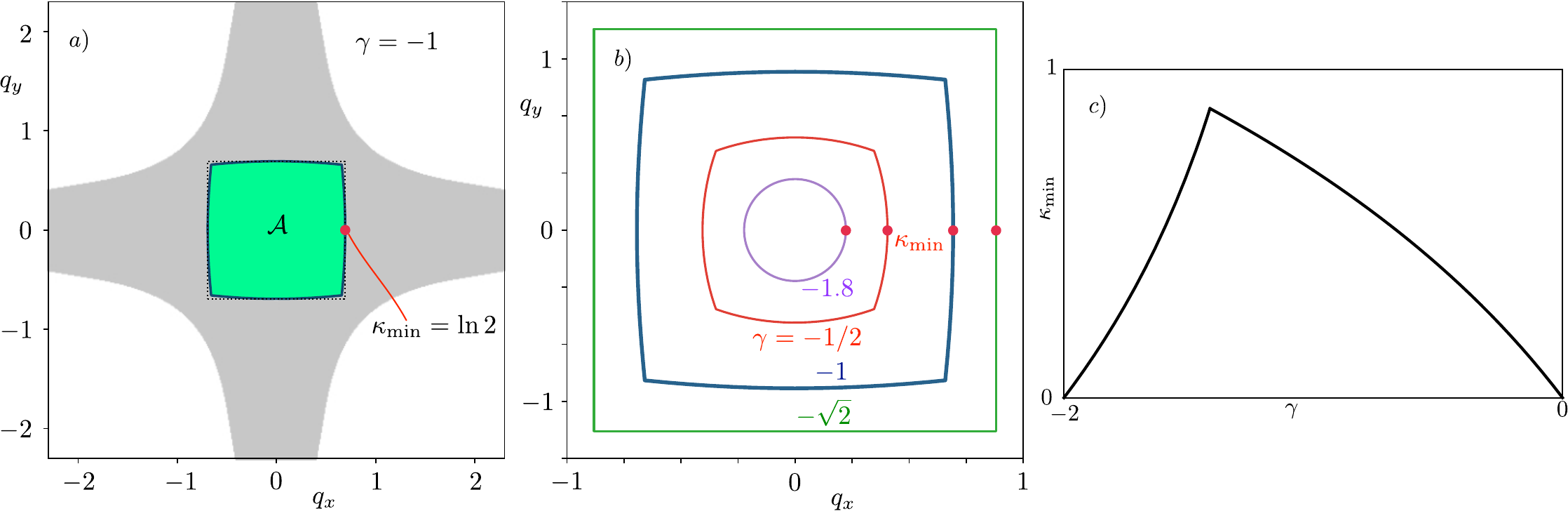}}
\caption{Panel a): Amoeba plot for the function $D(\bm{k}+i\bm{q})$ that governs the exponential decay of $U_{\text{QWZ}}(\bm{r})$. The grey area contains the points $\bm{q}$ at which $D$ vanishes for some $\bm{k}$, computed for $\gamma=-1$. The complementary region (green) is the analyticity domain ${\cal A}$. Points on the boundary of ${\cal A}$ give the decay rate in a particular direction, the slowest decay rate $\kappa_{\rm min}$ is the boundary point closest to the origin (red dot).
Panel b): Analyticity domain ${\cal A}$ for several values of $\gamma$. It is convex, interpolating between a disc and a square of size $2\kappa_{\rm min}\times 2\kappa_{\rm min}$.
Panel c): $\gamma$-dependence of the slowest decay rate $\kappa_{\rm min}$, according to Eq.\ \eqref{kappaclosed}.  The decay rate goes to zero as $\gamma$ approaches the end points of the interval $(-2,0)$. The decay rate peaks at $\kappa_{\rm min}= 0.88$ for $\gamma=-\sqrt 2$.
}
\label{fig_amoeba}
\end{figure*}

\subsection{Exponential locality}
\label{sec_locality}

The exponential decay of the Floquet operator on the lattice follows from analyticity of $U_{\text{QWZ}}$. Let us work this out and calculate the decay rate. We define the Fourier transform
\begin{equation}
U_{\text{QWZ}}(\bm{r})=(2\pi)^{-2}\int_{\mathbb{T}^2}d\bm{k}\,U_{\rm QWZ}(\bm{k})e^{-i\bm{k}\cdot\bm{r}},\label{Urdef}
\end{equation}
with $\mathbb{T}^2$ the Brillouin zone torus and $\bm{r}\in\mathbb{Z}^2$ a site on the 2D lattice.

For $-2<\gamma<0$ the function $U_{\text{QWZ}}(\bm{k})$ is analytic for real momenta and it can be analytically continued to complex momenta $\bm{k}+i\bm{q}$ with a sufficiently small imaginary part $\bm{q}\in {\cal A}$. The domain ${\cal A}$ of analyticity is bounded by the roots of $D(\bm{k}+i\bm{q})$, which are the only singularities in the complex plane. The so-called \cite{Mik01,Wan24} ``amoeba plot'' in Fig.\ \ref{fig_amoeba}a shows the $\bm{q}$-region for which $D(\bm{k}+i\bm{q})=0$ for some real $\bm{k}$. The complementary region containing the origin is the analyticity domain ${\cal A}$.

A shift of the integration contour from $\mathbb{T}^2$ to $\mathbb{T}^2-i\bm{q}$ does not change the integral if $\bm{q}\in{\cal A}$. This shift introduces an exponential factor $e^{-\bm{q}\cdot\bm{r}}$ in the Fourier transform \eqref{Urdef}, producing the upper bound
\begin{equation}
\| U_{\text{QWZ}}(\bm{r})\|\leq f(\bm{q})e^{-\bm{q}\cdot\bm{r}},\label{Ubound}
\end{equation}
for any matrix norm $\|U\|$. Since the roots of $D$ come in complex conjugate pairs, one can always choose the sign such that $\bm{q}\cdot\bm{r}>0$. The exponential decay rate in the direction $\hat{\bm{r}}=\bm{r}/|\bm{r}|$ is therefore bounded from below by the support function of ${\cal A}$,
\begin{equation}
\kappa(\bm{r})=\max_{\bm{q}\in{\cal A}}\bm{q}\cdot\hat{\bm{r}}.
\end{equation}

Because ${\cal A}$ is convex (as it must be for any amoeba \cite{Mik01}), the point $\bm{q}_{0}$ on the boundary of ${\cal A}$ that is closest to the origin gives the slowest decay rate $\kappa_{\rm min}=|\bm{q}_{0}|$ (slowest among all the directions). Algebraically,
\begin{equation}
\kappa_{\rm min}=\min\bigl\{\bigl|\ln(1-\gamma)\bigr|,\;\bigl|\ln|1+\gamma|\bigr|\bigr\},
\label{kappaclosed}
\end{equation}
for $\bm{q}_0$ aligned with the $x$ or $y$-axes, and we have checked this closed-form expression by numerical evaluation of the Fourier transform \cite{note5}.

What we have described is the generic mechanism by which the spectral projector of a gapped local Hamiltonian acquires exponentially decaying matrix elements \cite{Kim25}. A nonzero Chern number obstructs the existence of an exponentially localized Wannier basis \cite{Tho84,Bro07}, but not the exponential decay of the projector kernel itself --- which is all that the quantum walk requires.

\subsection{Robustness of the Dirac cone}
\label{sec_QWZrobustness}

\begin{figure}[tb]
\centerline{\includegraphics[width=0.8\linewidth]{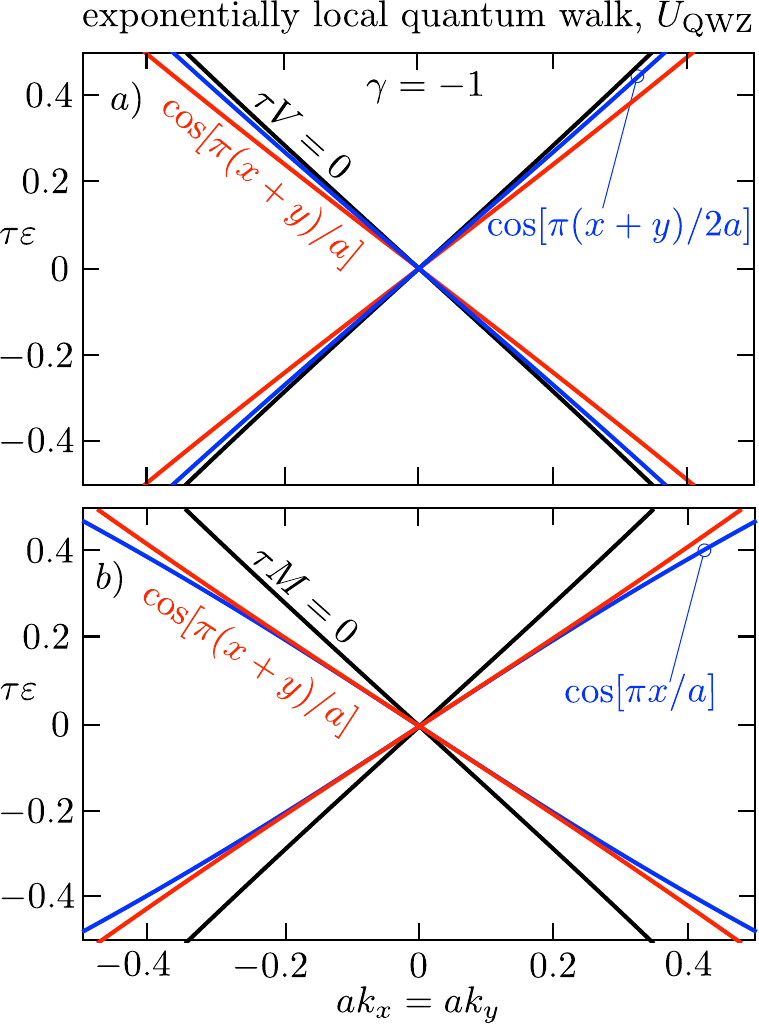}}
\caption{Same as Figs.\ \ref{fig_squared} and \ref{fig_HC}, but for the exponentially local quantum walk with Floquet operator \eqref{UQWZ} (at parameter $\gamma=-1$). Neither the scalar potential (panel a) nor the in-plane magnetization (panel b) gaps the Dirac point.
}
\label{fig_QWZ_potential}
\end{figure}

We have tested the robustness of the Dirac cone in the same way as in Sec.\ \ref{sec_stability}, by introducing either a cosine scalar potential $V(\bm{r})=V\cos(\bm{Q}\cdot\bm{r})$ or an in-plane magnetization $\bm{M}(\bm{r})=(M,M,0)\cos(\bm{Q}\cdot\bm{r})$. The band structure is computed by the momentum-ring construction of App.\ \ref{app_bandstructure}, unchanged except that the $2\times 2$ blocks are now given by Eq.\ \eqref{UQWZ}. We expect that the Dirac point remains gapless, protected by the symmorphic chiral or time-reversal symmetry, and that is indeed what we find (see Fig.\ \ref{fig_QWZ_potential}).

\section{Conclusion}

Unpaired Dirac cones are known to appear in the quasi-energy spectrum of periodically driven (Floquet) lattices \cite{Ley16,Liu20}. The periodicity of the quasi-energy invalidates the counting arguments behind the Nielsen-Ninomiya fermion-doubling no-go theorem \cite{Nie81}, and Floquet systems permit an unpaired cone at quasi-energy $\varepsilon=0$ by pushing its partner to the upper bound $\varepsilon=\pi/\tau$ of the Brillouin zone \cite{Bes21,Iad24,Bri25,Bak25,Bak26}. 

What has remained open, and what we have addressed here, is whether an unpaired $\varepsilon=0$ cone obtained in this way from a 2D quantum walk has the same topological protection as the Dirac cone on the 2D surface of a topological insulator. Our answer is negative: for a spin-1/2 walker the protecting symmorphic symmetries (chirality, time-reversal) cannot coexist with locality in two dimensions --- and the surviving non-symmorphic symmetries are broken by spatial inhomogeneities.

The contrast with one dimension is instructive. In 1D the quasi-energy periodicity genuinely reconciles locality, symmorphic symmetry, and a single Dirac cone \cite{Ced18}: Feynman's checkerboard walk $U=e^{-iap_x\sigma_x}$ realizes all three. Our results show that this route does not carry over to a 2D spin-1/2 walker: the doubler can still be displaced to $\varepsilon=\pi/\tau$ --- but the price is the loss of the symmorphic symmetry, and with it the loss of topological protection.

This conclusion is directly relevant for existing experimental platforms. Unpaired Dirac cones in Floquet band structures have been realized in photonic lattices \cite{Ley16,Liu20}, and the Ho-Chalker quantum walk is the momentum-space description of oriented scattering networks implemented in coupled optical resonators \cite{Lia13,Pas14,Del17}. Our analysis predicts that these unpaired cones, unlike their topological-insulator counterparts, are gapped by short-range correlated disorder, with a gap that grows quadratically in the disorder amplitude.

While we considered a two-band spin-1/2 implementation, it would be of interest to investigate whether additional ``flavor'' degrees of freedom, proposed recently as a way to avoid fermion doubling \cite{Bak25}, could provide for the topological protection that two bands lack. The alternative route, which we have followed in Sec.\ \ref{sec_exploc}, is to give up strict locality of the Floquet operator. Our no-go argument then loses its force, and it does so at the earliest possible point: \textit{Exponentially decaying} couplings already suffice to reconcile a single cone with symmorphic chiral and time-reversal symmetry.

\acknowledgments

We used several AI models (Claude Opus/Fable, GPT Pro) as an interactive tool to explore the topic.\\
This work was supported by the Netherlands Organisation for Scientific Research (NWO/OCW), as part of Quantum Limits (project number {\sc summit}.1.1016).\\
Data sets are available at a \href{https://doi.org/10.5281/zenodo.21890212}{Zenodo repository}.

\appendix

\section{Bandstructure calculations}
\label{app_bandstructure}

We outline the calculation that gives the results presented in Sec.\ \ref{sec_stability}. A similar calculation in the context of tangent fermions is in Ref.\ \onlinecite{Don22}.

We consider the effect of potential scattering on the 2D quantum walks considered in the main text. These have free evolution operators given by Eqs.\ \eqref{FXYYX}, \eqref{FXYXY}, and \eqref{UHC} upon substitution of $p_\alpha\mapsto k_\alpha$,
\begin{subequations}
\begin{align}
&U_{\sqrt{XYYX}}(\bm{k})=e^{-ik_x\sigma_x/2}e^{-ik_y\sigma_y}e^{-ik_x\sigma_x/2},\\
&U_{\sqrt{XYXY}}(\bm{k})=\bigl(e^{-ik_x\sigma_x/2}e^{-ik_y\sigma_y/2}\bigr)^2,\\
&U_{\text{Ho-Chalker}}(\bm{k})=e^{-i(k_x+k_y)\sigma_x/2}e^{-i(k_x-k_y)\sigma_y/2}.
\end{align}
\end{subequations}
(We set $a$ equal to unity.)

The cosine potential $V(\bm{r})=V\cos(\bm{Q}\cdot\bm{r})$ couples $\bm{k}$ to $\mathbf k\pm\mathbf Q$, with amplitude $V/2$. For a commensurate wavevector the momenta
\begin{equation}
\bm{k}_j=\bm{k}+j\,\bm{Q} ,\;\; j=0,1,\dots,N-1 ,
\end{equation}
close into a ring after $N$ steps, with $N$ the smallest integer for which $N\mathbf Q$ is a reciprocal lattice vector. 

The quasi-energies $e^{-i\tau\varepsilon_n}$ are the eigenvalues of the $2N\times 2N$ matrix
\begin{equation}
{\cal U}={\cal V}\begin{pmatrix}
U(\bm{k})&&&&\\
&U(\bm{k}+\bm{Q})&&&\\
&&\ddots&&\\
&&&U(\bm{k}+(N-1)\bm{Q})
\end{pmatrix}{\cal V}.\label{Ucheckerboard}
\end{equation}
The $2\times 2$ blocks are coupled by the matrix
\begin{equation}
{\cal V}=\exp(-\tfrac{1}{4}iV\tau A\otimes\sigma_0),
\end{equation}
acting as the identity on each block. The $N\times N$ tri-diagonal matrix $A$, with elements
\begin{equation}
  A_{jl}=\delta_{l,\,j+1}+\delta_{l,\,j-1}\;\;(\text{indices mod }N),
 \end{equation}
 shifts $\bm{k}_j$ to $\bm{k}_{j\pm 1}$. 
 
 Two examples (with $u\equiv V\tau/4$) \cite{note4}:
 \begin{align}
  &N=2:\;\;A=\begin{pmatrix}0&2\\2&0\end{pmatrix}\nonumber\\
  &\Rightarrow
  {\cal V}=\begin{pmatrix}
\cos 2u&-i\sin 2u\\
-i\sin 2u&\cos 2u
\end{pmatrix}\otimes\sigma_0,\label{Nistwo}
\end{align}
\begin{align}
&N=4:\;\;A=\begin{pmatrix}
 0 & 1 & 0 & 1 \\
 1 & 0 & 1 & 0 \\
 0 & 1 & 0 & 1 \\
 1 & 0 & 1 & 0
\end{pmatrix}\Rightarrow\\
& {\cal V}=\begin{pmatrix}
 \cos^2 u & -\frac{1}{2} i \sin 2u & -\sin^2 u & -\frac{1}{2} i \sin 2u \\
 -\frac{1}{2} i \sin 2u & \cos^2 u & -\frac{1}{2} i \sin 2u & -\sin^2 u \\
 -\sin^2 u & -\frac{1}{2} i \sin 2u & \cos^2 u & -\frac{1}{2} i \sin 2u \\
 -\frac{1}{2} i \sin 2u & -\sin^2 u & -\frac{1}{2} i \sin 2u & \cos^2 u 
 \end{pmatrix}\otimes\sigma_0.
\end{align}

For $N=2$ and $\bm{k}=0$ we may work in a basis where $U$ is diagonal,
\begin{equation}
U=\operatorname{diag}(1,1,e^{i\phi},e^{-i\phi}),\label{Uphi}
\end{equation}
where $\phi=\tau\varepsilon(\bm{Q})$ at $V=0$. The simple closed form \eqref{Egap} then results for the gap at the Dirac point. The gapless case $\phi=\pi$ has a simple interpretation: it corresponds to $U(\bm{Q})=-\sigma_0$, so Eq.\ \eqref{Uphi} becomes $U=\operatorname{diag}(1,1,-1,-1)$, which anticommutes with $A\otimes\sigma_0$. Hence ${\cal V}U{\cal V}=U{\cal V}^{-1}{\cal V}=U$ and the staggered potential cancels to all orders in $V$.

\section{Parity of a Laurent polynomial map}
\label{app_parity}

We give a proof that the parity ${\cal P}$ defined in Eq.\ \eqref{Pdef} must be equal to +1, to exclude an unpaired Dirac cone at $\varepsilon=0$ in the case of a local 2D   Dirac quantum walk with time-reversal symmetry.

The condition of a local Floquet operator $U(\bm{k})$ can be expressed algebraically by demanding that the functions $n_j(k_x,k_y)$ in the parameterization \eqref{UTRS} are finite Laurent polynomials in $u=e^{iak_x}$ and $w=e^{iak_y}$,
\begin{equation}
n_j(u,w)=\sum_{m,n=-N}^N c^{(j)}_{mn} u^m w^n,\;\;c^{(j)}_{mn}\in\mathbb{C}.
\end{equation}
The integer $N$ gives the range of sites on the lattice that are coupled by $U$.

Since $n_j$ is real-valued on $\mathbb{T}^2$, we must have
\begin{equation}
c^{(j)}_{mn}=\left(c^{(j)}_{-m,-n}\right)^\ast.\label{realvalued}
\end{equation}
Moreover, since $n_0$ is an even function of $\bm{k}$ and $n_x,n_y,n_z$ are odd, 
\begin{equation}
c^{(0)}_{mn}=c^{(0)}_{-m,-n},\;\;c^{(\alpha)}_{mn}=-c^{(\alpha)}_{-m,-n},\;\;\alpha\in\{x,y,z\},
\end{equation}
hence $c^{(0)}_{mn}$ is real while $c^{(x)}_{mn},c^{(y)}_{mn},c^{(z)}_{mn}$ are purely imaginary.

The four functions $n_0$, $\tilde{n}_\alpha=in_\alpha$, $\alpha\in\{x,y,z\}$ are finite Laurent polynomials in $u,w$ with real coefficients. The unitarity constraint \eqref{TRSunitarity} reads
\begin{equation}
{n}_0^2(u,w)-\tilde{n}_x^2(u,w)-\tilde{n}_y^2(u,w)-\tilde{n}_z^2(u,w)-1=0,\label{tilden}
\end{equation}
on the torus, $|u|=1=|w|$. The left-hand-side is a finite Fourier series in $k_x,k_y$; the uniqueness of the Fourier expansion then implies that all Fourier coefficients vanish. Hence the left-hand-side is identically zero for all nonzero $u,w\in\mathbb{C}$.

If we now consider real $u,w\in\mathbb{R}^\ast$ (real $\neq 0$), the functions $n_0,\tilde{n}_x,\tilde{n}_y,\tilde{n}_z$ are real and Eq.\ \eqref{tilden} tells us that
\begin{align}
&{n}_0^2(u,w)=1+\tilde{n}_x^2(u,w)+\tilde{n}_y^2(u,w)+\tilde{n}_z^2(u,w)\nonumber\\
&\Rightarrow {n}_0^2(u,w)\geq 1,\;\;u,w\in\mathbb{R}^\ast.\label{tildennew}
\end{align}
Here we make essential use of locality: a non-local Floquet operator can be represented by an \textit{infinite} Laurent polynomial, but then the extension of the identity \eqref{tilden} to real $u,w$ fails.

The time-reversally invariant momenta correspond to the four points $(u,w)=(1,1),(-1,1),(1,-1),(-1,-1)$. The parity \eqref{Pdef} is given by
\begin{equation}
{\cal P}=\prod_{u,w\in\{+1,-1\}}\operatorname{sign}{n}_0(u,w).\label{Papp}
\end{equation}
We now proceed to show that ${\cal P}=+1$.

Since, in view of Eq.\ \eqref{tildennew}, $n_0(u,w)\neq 0$ for $u,w\in\mathbb{R}^\ast$, it has a definite sign in each of the four quadrants of the $u,w$ plane. We can determine the sign from the asymptotic behavior of $n_0(u,w)$, along curves that isolate a single dominating monomial.

We parameterize
\begin{equation}
(u,w)=(\epsilon R^s,\eta R),\;\;R>0,\;\;s>2N.
\end{equation}
Along this trajectory one has
\begin{equation}
n_0(\epsilon R^s,\eta R)=\sum_{m,n=-N}^N c_{mn}^{(0)} \epsilon^m \eta^n R^{sm+n}.\label{n0R}
\end{equation}
Since $s>2N$, the scaling exponent $sm+n$ satisfies $m_1>m_2\Rightarrow sm_1+n_1>sm_2+n_2$, hence for $R\rightarrow\infty$ the Laurent polynomial \eqref{n0R} is uniquely dominated by the monomial corresponding to the lexicographically largest exponent pair $M,K$ for which $c_{MK}^{(0)}\neq 0$:
\begin{equation}
\begin{split}
&M=\max\{m:c^{(0)}_{mn}\neq 0 \;\;\text{for some}\;\; n\},\\
&K=\max\{n:c^{(0)}_{Mn}\neq 0\}.
\end{split}
\end{equation}

The asymptotic behavior
\begin{equation}
\lim_{R\rightarrow\infty} R^{-sM-K}n_0(\epsilon R^s,\eta R)=c^{(0)}_{MK}\epsilon^M\eta^K
\end{equation}
allows to evaluate the parity \eqref{Papp},
\begin{equation}
{\cal P}=\prod_{\epsilon,\eta\in\{+1,-1\}}[\operatorname{sign}c^{(0)}_{MK} ]\epsilon^M \eta^K=(-1)^{2M}(-1)^{2K}=+1.
\end{equation}

This proof also shows how the 1D case is essentially different: there is then only a single variable $u$, and a single element $\epsilon$, so the parity ${\cal P}=\prod_{\epsilon\in\{+1,-1\}} \epsilon^M=(-1)^M$ can be $-1$ for $M$ odd.

\section{Transfer matrix of the Ho-Chalker quantum walk}
\label{app_Klein}

To perform the mode-matching calculation of the Ho-Chalker quantum walk in the Klein tunneling geometry of Sec.\ \ref{sec_Klein}, we need the transfer matrix ${\cal T}(x)$, relating the scattering state at $x$ and $x+1$. We compute that here, for a general potential profile $V(x)$.

The scattering state $\psi(x)$ at quasi-energy $\varepsilon$ is an eigenstate with eigenvalue $e^{-i\varepsilon\tau}$ of the Ho-Chalker Floquet operator \eqref{FHC}, evaluated at $p_y=0$: 
\begin{equation}
\begin{split}
&F=e^{-iV(x)\tau/2}U_0 e^{-iV(x)\tau/2},\\
&U_0= e^{-iap_x\sigma_x/2}e^{-iap_x\sigma_y/2}.
\end{split}
\end{equation}
If we define $\phi(x)=e^{-iV(x)\tau/2}\psi(x)$, we have
\begin{align}
(U_0\phi)(x)=\lambda_{x}\phi(x),\;\;\lambda_{x}=e^{-i\varepsilon\tau}e^{iV(x)\tau},
\end{align}
which amounts to a three-term recursion relation,
\begin{subequations}
\label{recursioneq}
\begin{align}
&M_-\phi(x+1)+M_0\phi(x)+M_+\phi(x-1)=\lambda_{x}\phi(x),\\
&M_\pm=\kappa|u_\pm\rangle\langle v_\pm|,\;\;M_0=1-M_+-M_-,
\end{align}
\end{subequations}
given in terms of eigenstates $|u_\pm\rangle$ and $|v_\pm\rangle$ of $\sigma_x$ and $\sigma_y$, respectively,
\begin{equation}
\begin{split}
&|u_\pm\rangle=2^{-1/2}{{1}\choose{\pm 1}},\;\;|v_\pm\rangle=2^{-1/2}{{1}\choose{\pm i}},\\
&\kappa=\langle u_\pm|v_\pm\rangle=\tfrac{1}{2}(1+i).
\end{split}
\end{equation}

By projecting onto $\langle u_\pm|$ we can cast the three-term recursion \eqref{recursioneq} into a pair of two-term recursions,
\begin{subequations}
\begin{align}
&\begin{cases}
\langle u_-|M_-|x+1\rangle+\langle u_-|M_0|x\rangle=\lambda_{x}\langle u_-|x\rangle,\\
\langle u_+|M_0|x\rangle+\langle u_+|M_+|x-1\rangle=\lambda_{x}\langle u_+|x\rangle,
\end{cases}\\
&\Rightarrow\begin{cases}
\kappa \langle v_-|x+1\rangle=[\lambda_{x}-1]\langle u_-|x\rangle+\kappa\langle v_-|x\rangle,\\
\kappa \langle v_+|x\rangle=[\lambda_{x+1}-1]\langle u_+|x+1\rangle+\kappa\langle v_+|x+1\rangle,
\end{cases}
\end{align}
\end{subequations}
where we denote $\phi(x)\equiv|x\rangle$.

The corresponding transfer matrix ${\cal T}(x)$, relating
\begin{equation}
\phi(x+1)={\cal T}(x)\phi(x),
\end{equation}
is given by
\begin{widetext}
\begin{align}
&{\cal T}(x)=\frac{1}{2\lambda_{x+1}}\begin{pmatrix}
2+2 \lambda_{x} \lambda_{x+1}-(1+i)(\lambda_{x}-i\lambda_{x+1})&-2i-2\lambda_{x}\lambda_{x+1}+(1+i)(\lambda_{x}+\lambda_{x+1})\\
2i-2\lambda_{x}\lambda_{x+1}+(1-i)(\lambda_{x}+\lambda_{x+1})&2+2 \lambda_{x} \lambda_{x+1}-(1-i)(\lambda_{x}+i\lambda_{x+1})
\end{pmatrix}.
\end{align}

For the mode matching calculation of Sec.\ \ref{sec_Klein} we need ${\cal T}(0)$ and ${\cal T}(w)$ for $\lambda_0=0=\lambda_{w+1}$, $\lambda_1=e^{iV_0\tau}=\lambda_w$, when
\begin{subequations}
\begin{align}
&{\cal T}(0)=\tfrac{1}{2}e^{-iV_0\tau}\begin{pmatrix}
(1+i)(-i+e^{iV_0\tau})&(1-i)(1-e^{iV_0\tau})\\
(1+i)(1-e^{iV_0\tau})&(1-i)(i+e^{iV_0\tau})
\end{pmatrix},\\
&{\cal T}(w)=\tfrac{1}{2}\begin{pmatrix}
(1-i)(i+e^{iV_0\tau})&(1-i)(1-e^{iV_0\tau})\\
(1+i)(1-e^{iV_0\tau})&(1+i)(-i+e^{iV_0\tau})
\end{pmatrix}.
\end{align}'
\end{subequations}
\end{widetext}

We solve
\begin{equation}
\psi(1)={\cal T}(0)\psi(0),\;\psi(w+1)={\cal T}(w)\psi(w),
\end{equation}
with $\psi(x)$ given by Eqs.\ \eqref{phioutsidebarrier} and \eqref{phiinsidebarrier}, for the coefficients $t,r,A_+,A_-$. The resulting transmission probability $T=|t|^2$ is given by Eq.\ \eqref{Tresult}.


\begin{thebibliography}{99}
\bibitem{note1} Because of the $\pm\varepsilon$ (chiral) symmetry of the quasi-energy spectrum, the dispersion relation follows directly from $\cos\tau\varepsilon=\tfrac{1}{2}\operatorname{Tr}U$, upon substitution of $T_\alpha\mapsto e^{iak_\alpha}$.
\bibitem{note3} For consistency with the choice of Pauli matrices in $H_{\rm Dirac}$, we have rotated $\sigma_z\mapsto\sigma_y$ in the Ho-Chalker unitary \eqref{UHC}.
\bibitem{note4} The factor of two in Eq.\ \eqref{Nistwo} was missed in Ref.\ \onlinecite{Don22}. As a consequence, the band structures for the 1D and 2D cosine potentials in that paper are calculated and plotted for $(V/2) \cos \pi x$ and $(V/2) \cos(\pi x+\pi y)$, not as stated for $V \cos \pi x$ and $V \cos(\pi x+\pi y)$.
\bibitem{note5} The upper bound\ \eqref{Ubound} allows for a more rapid decay than $\kappa(\bm{r})$, but not a more slower decay. Numerically we have found that $\kappa(\bm{r})$ saturates the bound for $\bm{r}$ along the $x$- or $y$-axes. The decay along the diagonal is more rapid.
\end{thebibliography}
\end{document}